\documentclass[reprint,superscriptaddress,a4paper,aps,showpacs,longbibliography]{revtex4-2}
\usepackage{graphicx}
\graphicspath{{./},{/home/user/fig/publi/muon/}}
\usepackage{amsfonts}
\usepackage{amsbsy}
\usepackage{amsmath}
\usepackage{diagbox}

\begin{document}

\title{Power-law dynamics in the spin-liquid kagome lattices SrCr$_{8}$Ga$_{4}$O$_{19}$ and ZnCu$_3$(OH)$_6$Cl$_2$}
\author{P. Dalmas de R\'eotier}
\affiliation{Universit\'e Grenoble Alpes, CEA, Grenoble INP, IRIG-PHELIQS, F-38000 Grenoble, France}
\author{A. Yaouanc}
\affiliation{Universit\'e Grenoble Alpes, CEA, Grenoble INP, IRIG-PHELIQS, F-38000 Grenoble, France}
\date{\today}

\begin{abstract}

  We consider the polarization function $P^{\rm exp}_Z(t)$ measured by the muon-spin relaxation ($\mu$SR) technique for the SrCr$_{8}$Ga$_{4}$O$_{19}$ and ZnCu$_3$(OH)$_6$Cl$_2$ spin-liquid systems. We show the functional form of $P^{\rm exp}_Z(t)$ to imply that, in the temperature range of order 0.1~K, the spectral-weight function $F(\omega)$ of the magnetic correlations scales with $1/|\omega|^{1 -x}$ ($0 < x < 1$) in the energy range of one microelectronvolt, i.e.\ $\hbar \omega \approx 1\ \mu$eV. We derive  the parameters involved in $F(\omega)$ from fits to available experimental data. Inelastic neutron scattering data probing $F(\omega)$ in the millielectronvolt energy range are consistent with a more conventional behavior. These differences could be due to a variety of spin-dynamics mechanisms, i.e.\ intrinsic to the kagome layer or related to the magnetic defects that have been evidenced in these compounds, acting at different energies. New $\mu$SR measurements are proposed and theoretical developments are suggested to pinpoint the mechanisms at play.

\end{abstract}

\maketitle

{\sl Introduction.}
The spin liquid (SL) state is an exotic phase of magnetic materials which remain magnetically disordered  down to the lowest accessible temperatures despite the presence of strong interactions. It does not break symmetries and it hosts long-range entanglement, fractional excitations and an emerging gauge-field. It is believed that the interplay of quantum fluctuations and geometric frustration favor this SL state \cite{Balents10,Savary17,Zhou17,Broholm20}. Systems with magnetic moments sitting on a kagome lattice, i.e. the archetypal geometrically frustrated lattice consisting of corner-sharing triangles, are thus prone to display SL states at low temperature.

The SrCr$_{8}$Ga$_{4}$O$_{19}$ (SCGO) compound, in which Cr$^{3+}$ $S = 3/2$ spins lie on the vertices of the pyrochlore-slab (kagome bilayer) lattice \cite{Obradors88,Ramirez90,Broholm90}, was the first  discovered inorganic SL material \footnote{The absence of spontaneous oscillations in $\mu$SR zero-field experiments down to 50~mK \cite{Uemura94} was taken as a support to this conclusion. Yet, as it has been recognized later on \cite{Dalmas06}, a compound which displays magnetic Bragg reflections may not exhibit muon spin spontaneous precession. However, specific heat and neutron diffraction data support the existence of the SL ground state in SCGO \cite{Ramirez90,Broholm90}}. 
Since the effect of quantum fluctuations is dramatically magnified by small spin values, it is desirable to find compounds with spins 1/2 located on the vertices of a kagome lattice. The kagome Heisenberg antiferromagnet ZnCu$_3$(OH)$_6$Cl$_2$ called herbertsmithite (HS) with Cu$^{2+}$ ions fulfils that purpose and was the first reported example of such a material \cite{Shores05,Helton07,Norman16}. It consists in a single kagome plane in contrast to SCGO, and moreover it can be prepared as fairly large single crystals \cite{Chu11,Han12}. The broad continuum of spin excitations typical of spinons, which is  detected by inelastic neutron scattering (INS) \cite{Han12}, is a striking evidence for the SL ground state. It is of gapless nature according to nuclear magnetic resonance (NMR) \cite{Khuntia20}, in agreement with prior experimental results, e.g., specific heat \cite{Helton07}, but at variance from other studies that are consistent with a gap of order 1~meV \cite{Fu15,Han16}.

Measurements performed for a powder sample of SCGO in zero-field (ZF) by the positive muon-spin relaxation ($\mu$SR) technique showed that the longitudinal polarization function, which monitors the evolution with time $t$ of the polarization of implanted muons, decays as an exponential-power-law \cite{Uemura94},
\begin{eqnarray}
P_Z^{\rm exp}(t) = \exp \left [ - \left (\frac {t}{T_1} \right )^\beta  \right ],
\label{intro_1}
\end{eqnarray}
with $ 1.5 \leq \beta  < 2.0 $ below 2~K.  The value of $\beta$ is clearly smaller than $2$ \cite{Keren00}. Hence, $P_Z^{\rm exp}(t)$ is neither an exponential nor a Gaussian function, and is therefore not compatible with the prediction of the conventional theory \cite{Hayano79}. Furthermore, the magnetic field response is in stark contrast to expectation \cite{Uemura94}. Indeed, the close-to Gaussian lineshape of the spectra suggests the presence of a distribution of quasi-static fields at the probe site. Now, the applied field magnitude at which the muon spin is decoupled from these internal fields is much higher than expected from the distribution width estimated from the ZF spectrum. Qualitatively similar results have been obtained for HS \cite{Mendels07}. This suggests the same relaxation mechanism to be at play in SCGO and HS.

Relying on a key information extracted from INS \cite{Broholm20}, we propose an explanation of the functional form of the measured relaxation. It is found to reflect the power-law decay of the spectral-weight function of the magnetic fluctuations in the microelectronvolt range \footnote{Note that power-law spectral densities have already been derived from the field-dependence of spectra recorded in spin liquids and analysed with Eq.~\ref{intro_1}, see e.g.\ Refs.~\onlinecite{Keren04,Kermarrec11}. The present study provides a general expression for $P_Z(t)$ which may deviate sizably from an exponential-power-law function under field.}. From this result, we discuss the energy dependence of the electronic spin correlation function.

{\sl The $\mu$SR longitudinal polarization function $P_Z(t)$.} 
Following the methodology introduced for NMR --- see Ref.~\onlinecite{Kubo54a} --- the polarization function $P_Z(t)$ can be computed in terms of  correlation functions of the magnetic field at the muon site \cite{McMullen78,Dalmas92},
\begin{eqnarray}
  P_Z(t) & = & \exp \left [ - \Psi_Z(t) \right ],
\label{Th_1}
\end{eqnarray}
where the Cartesian $Z$ axis is set along the initial muon spin polarization and
\begin{eqnarray}
  \Psi_Z(t)   \label{Th_2}
  & = & 2\pi\gamma_\mu^2 \int_0^t
  \left (t - \tau \right ) \cos(\omega_\mu \tau)[ \Phi_{XX}(\tau) +  \Phi_{YY}(\tau)]
  {\rm d} \tau .\cr & &
\end{eqnarray}
A second term has been omitted on the right-hand side of Eq.~\ref{Th_2} since it is negligibly small in the fast dynamics limit we are considering \cite{Yaouanc11}. We have defined the muon spin precession angular frequency $\omega_\mu = \gamma_\mu B_{\rm ext}$ where $B_{\rm ext}$ is the intensity of the external magnetic field applied along the $Z$ axis and $\gamma_\mu = 8.51616 \times 10^8 \, {\rm rad \, s}^{-1} {\rm T}^{-1}$ is the muon gyromagnetic ratio. We have introduced  the field correlation function $\Phi_{\alpha \alpha}(\tau)$ for Cartesian component $B_\alpha$ of the local magnetic field at the muon site.

Since the magnetic ions are located at crystal positions, translational symmetry matters. That property is best exploited by expressing $\Phi_{\alpha \alpha}(\tau)$ in reciprocal space \cite{Moriya62}. Then \footnote{See, e.g., Chap.~10 in Ref.~\onlinecite{Yaouanc11}},
\begin{eqnarray}
 & &  \Psi_Z(t)   =  \label{Th_3} \\ 
 & &  {\mathcal D} \int_{v_c^*} \sum_{\gamma =X,Y,Z}{\mathcal A}^{\gamma \gamma}({\bf q}) \int_0^t
  \left (t - \tau \right ) \Lambda({\bf q},\tau)
  \cos(\omega_\mu \tau){\rm d} \tau \frac{{\rm d}^3 {\bf q}}{(2 \pi)^3}.\cr & &
\nonumber
\end{eqnarray}
Here ${\mathcal A}$ describes the muon-system coupling and ${\mathcal D}= (\mu_0/4\pi)^2\gamma_\mu^2(g \mu_{\rm B})^2/v_{\rm c}$, where $g$ is the spectroscopic factor and $v_{\rm c}$ is the volume per magnetic ion \footnote{The expression for the muon-system coupling is ${\mathcal A}^{\gamma\gamma^\prime}({\bf q}) = G^{X\gamma}({\bf q}) G^{\gamma^\prime X}(-{\bf q}) + G^{Y\gamma}({\bf q}) G^{\gamma^\prime Y}(-{\bf q})$ where $G^{\alpha\beta}({\bf q})$ are the components of the unitless tensor describing the coupling between the muon spin and the spin of the system under study. In the case of the SCGO and herbertsmithite, this is the dipolar interaction. See Ref.~\cite{Yaouanc11} for details about $G^{\alpha\beta}({\bf q})$.}. The first integration is over the first Brillouin zone. In Eq.~\ref{Th_3}, the spin correlation tensor $\Lambda({\bf q},t)$ with $\Lambda^{\alpha\beta}({\bf q},t) \equiv \langle S_\alpha({\bf q},t)S_\beta(-{\bf q},0)\rangle$, where ${\bf q}$ is a reciprocal space vector, is taken as scalar for simplicity. This assumption justifies the use of Eq.~\ref{Th_3} together with Eq.~\ref{Th_2} for the interpretation of data recorded on powders as we shall do. Should tensor $\Lambda({\bf q},t)$ not be scalar, the field correlations of Eq.~\ref{Th_2} are to be expressed in terms of correlations written in the crystal frame and possibly a polycrystalline average is to be performed. Still, the essence of the conclusions drawn in this paper would not be altered since they rest on the specific ${\bf q}$ and $t$ dependences of $\Lambda({\bf q},t)$ in the systems of interest here.

Rather than working with $\Lambda({\bf q},t)$, we consider its time Fourier transform $\Lambda({\bf q},\omega)$ to which we apply the fluctuation-dissipation theorem \cite{*[{See, e.g., }][{}] Lovesey86a,*Jensen91}) in the justified high-temperature limit ($k_{\rm B}T \gg \hbar\omega$), 
\begin{eqnarray}
\Lambda({\bf q},\omega) & = & \frac{2\pi v_c}{\mu_0 g^2\mu_{\rm B}^2} k_{\rm B}T \chi({\bf q})F({\bf q},\omega),
\label{Th_4}
\end{eqnarray}
where the scalar functions $\chi({\bf q})$ and $F({\bf q},\omega)$ are the wavevector dependent susceptibility and the spectral-weight function, respectively.

According to INS measurements performed on HS, $F({\bf q},\omega)$ can be taken independent of ${\bf q}$ \cite{Broholm20}. We shall assume this property to hold in general for SL kagome systems in the whole energy range. Hence the two integrals in Eq.~\ref{Th_3} can be separated, leading to
\begin{eqnarray}
  \Psi_Z(t) & = & 2 \gamma^2_\mu \Delta^2  t \int_{-\infty}^\infty F(\omega + \omega_\mu) f_t(\omega){\rm d}\omega,
\label{Th_5}
\end{eqnarray}
with
\begin{eqnarray}
  \Delta^2 &  = & \frac{\mu_0}{32\pi^2}k_{\rm B}T
  \int_{v_c^*} \sum_\gamma {\mathcal A}^{\gamma \gamma}({\bf q}) \chi({\bf q})
  \frac{{\rm d}^3{\bf q}}{(2\pi)^3}.
\label{Th_6}
\end{eqnarray}
Only $\Delta^2$ depends on the muon position characteristics, i.e.\ geometry and distance, in the expression of $\Psi_Z(t)$. An estimate can be obtained assuming the cross correlation of the spins at the source of the relaxation to be negligible. Then $\Delta^2$ represents the variance of the dipole-field distribution at the muon site taken to be Gaussian. It can be computed {\em \`a la} van Vleck \cite{vanVleck48,Yaouanc11}, as usual in $\mu$SR spectroscopy. The order of magnitude for $\Delta$ is in the range 0.1 to 1~T \footnote{This order of magnitude 0.1 to 1~T stems from the Gaussian field distribution root mean-square of order 0.1~mT commonly observed by $\mu$SR measurements in non-magnetic systems, after rescaling the magnetic moment at the origin of this field (of order one nuclear magneton) to the electronic moment, say 1 or 3\,$\mu_{\rm B}$.}. However we stress that Eq.~\ref{Th_6} surpasses the van Vleck formalism as it includes the effect of spatial correlations. These correlations are a hallmark of spin liquids and they can be measured in neutron scattering experiments. In Eq.~\ref{Th_5}, we have defined the auxiliary function 
\begin{eqnarray}
  f_t(\omega) = \frac{1-\cos(\omega t)}{\omega^2 t},
\label{Th_7}
\end{eqnarray}
which acts as a screening function selecting the $\omega$ range probed \cite{McMullen78}. It is of order $10^6$ to $10^8$~s$^{-1}$ for the time scale of a $\mu$SR experiment, and therefore $f_t(\omega)$ filters out modes at energies above $\approx 1~\mu$eV.

\begin{figure}
\includegraphics[width=\linewidth]{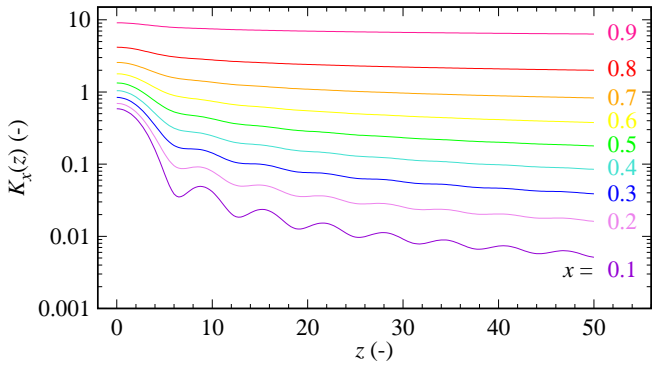}
\caption{Variation of $K_x(z)$ as a function of $z$ for different values of the exponent $x$ as indicated in the plot. Note the analytical expression $K_x(z = 0 ) = 1/(2-x)(1-x)$ \cite{Gradshteyn80}.
}
\label{K_x_z}
\end{figure}

\begin{figure}
\begin{picture}(255,290)
  \put(45,180){(a)}
  \put(0,150){\includegraphics[width=\linewidth]{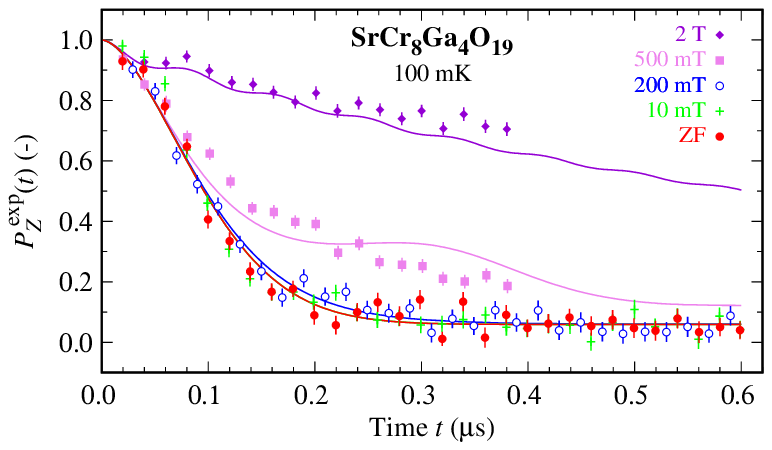}}
  \put(45,30){(b)}
  \put(0,0){\includegraphics[width=\linewidth]{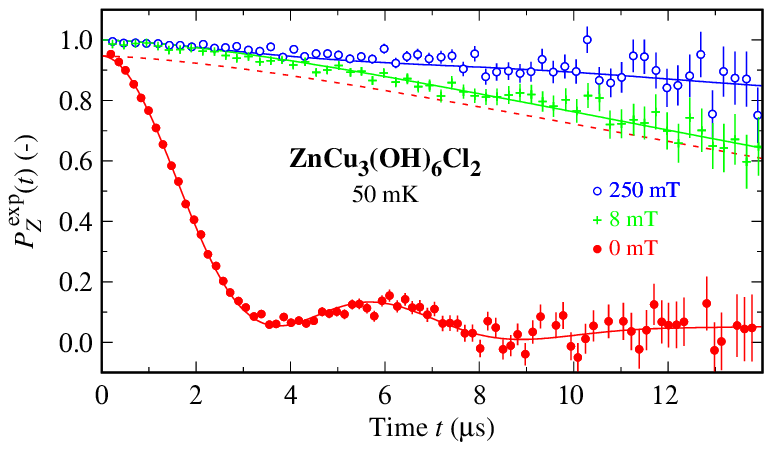}}
\end{picture}
\caption{Polarization functions measured for two kagome lattice systems for different applied fields as indicated in the graphs: (a) SCGO at 100~mK  and (b) HS at 50~mK. The data are respectively taken from Refs.~\onlinecite{Uemura94} and \onlinecite{Mendels07}. The lines are the result of a fit of Eq.~\ref{Th_1} together with Eq.~\ref{Th_11} to the respective combined sets of data with parameters given in Table \ref{table}. A small time-independent background is allowed in addition to the $P_Z(t)$ function, in particular for a better match of the low-field data \cite{Mendels07}. For the HS zero-field data, the effect of the coupling of the muon spin to nearby H, Cu and Cl nuclei is accounted for as in Ref.~\onlinecite{Mendels07} with similar parameters. The dotted line shows the model prediction at zero field, in the absence of this latter coupling. It implies the absence of resonance effect around 8~mT as found, e.g., in metallic copper \cite{{Abragam84,*Kreitzman86}}.
}
\label{data}
\end{figure}

At this juncture, we have to adopt a form for the spectral-weight function. The traditional choice is a Lorentzian function, corresponding to correlations exponentially decaying with time. Such a function \cite{Kubo54a,Hayano79} is associated with an electronic spin diffusion relaxation process \cite{vanHove54}. However, this mechanism can only be at play in the vicinity of the Brillouin zone center, while outside of it the correlations are expected to be Gaussian \cite{deGennes57,deGennes60}.
This dichotomy between the reciprocal space regions is at variance with the INS data which exhibit a ${\bf q}$-independent dynamics. Alternatively, a correlation function consistent with the local dynamics suggested by INS is a power-law function considered for spin-glass systems \cite{*[{See, e.g., }][] Ogielski85},
\begin{eqnarray}
  F(t) & = & F_x(t) = \frac{1}{2\pi}\left (\frac{\tau_{\rm e}}{t  + \tau_{\rm e}} \right )^x,
\label{Th_9}
\end{eqnarray}
where $0 < x < 1$ is an exponent \footnote{The $2\pi$ normalisation factor in Eq.~\ref{Th_9} stems from the definition $ F_x(t) = \int_{-\infty}^{\infty} F_x(\omega)\exp(-i\omega t)\,{\rm d}\omega/2\pi$ of the Fourier transform and the normalization $\int_{-\infty}^\infty F_x(\omega)\,{\rm d}\omega = 1$.}. Note that the distinctive feature for this algebraic function as compared with exponentially or Gaussian decaying functions is the absence of a time scale. Parameter $\tau_{\rm e}$ in Eq.~\ref{Th_9} should therefore be regarded as setting the scale for the correlations rather than their time dependence. Implicit in Eq.~\ref{Th_9} is the condition $t \gg \tau_{\rm e}$. Hence, the Fourier transform of $F_x(t)$ is 
\begin{eqnarray}
  F_x(\omega ) & = & \frac{ \tau^x_{\rm e} \sin(\pi x/2) \Gamma(1 -x)}{\pi |\omega|^{1-x}},
\label{Th_10}
\end{eqnarray}
where $\Gamma(x)$ is the gamma function \footnote{Strictly speaking, Eq.~\ref{Th_10} is the Fourier transform of $F(\tau) = (\tau_{\rm e}/\tau)^x/2\pi$. We have numerically checked that the neglect of $\tau_{\rm e}$ in the denominator of Eq.~\ref{Th_9} has no influence on the $P_Z(t)$ functions presented in Fig.~\ref{data} for the values derived for $\tau_{\rm e}$.}. With this expression we obtain
\begin{eqnarray}
 \Psi_Z(t) &  = & (\Upsilon f t)^{2 -x} K_x \left ( \omega_\mu ft \right ) ,
\label{Th_11}
\end{eqnarray}
with $\Upsilon^{2-x}$ = $2 \gamma^2_\mu \Delta^2 \tau_{\rm e}^x$ and 
\begin{eqnarray}
  K_x (z )& = & \frac{\sin(\pi x/2)\Gamma(1-x)}{\pi}
  \int_{-\infty}^\infty
\frac{1-\cos y}{|y + z |^{1-x}\, y^2} {\rm d} y. \cr & &
\label{Th_12}
\end{eqnarray}
Figure \ref{K_x_z} displays $K_x (z )$ for a selection of $x$ values.
For conventional systems the factor $f$ introduced in Eq.~\ref{Th_11} is equal to 1.
Here we acknowledge the possibility of formation in SLs of local and transient singlet spin pairs, resulting in a field acting for a fraction $0<f<1$ of the time on the muon, whereas this field is negligible for the remaining $1-f$ fraction. This {\em sporadic} relaxation mechanism was originally put forward in Ref.~\onlinecite{Uemura94}. The parameter $f$ could also reflect a diamagnetic response of the SL state believed to be of the $U(1)$-Dirac type, where the $U(1)$ gauge field is coupled to Dirac fermions \cite{Ran07}. In the following we do not discuss this point: we concentrate on the shape of $P_Z(t)$.

Equation \ref{Th_11} is our main analytical result, valid when $f t \gg \tau_{\rm e}$. As the numerical values derived from the fits will show, this condition is always fulfilled. The  $\Psi_Z(t)$ formula  is consistent with ZF results after setting $\beta = 2-x$ in Eq.~\ref{intro_1}. The field dependence of $P_Z(t)$ is solely determined by function $K_x (z )$, in which, remarkably, $\tau_{\rm e}$ does not appear. This is in line with the fact that $\tau_{\rm e}$ is not a spin correlation time, and in contrast to the usual descriptions of the field dependence of $P_Z(t)$ based either on the so-called dynamical Kubo-Toyabe model \cite{Hayano79} or on the Bloembergen-Purcell-Pound formula \cite{Bloembergen48} \footnote{In the two models the field evolution of $P_Z(t)$ depends on a correlation time. Note that both models converge in the fast-fluctuation limit \cite{Hayano79}.}. 

{\sl Analysis of published $P_Z^{\rm exp}(t)$.}
We now analyse published $P_Z^{\rm exp}(t)$ recorded at low temperature for SCGO and HS within the framework exposed above. The analysis, performed with a unique set of parameters for each of the systems, is rather successful as shown in Fig.~\ref{data}. We note that allowing a slight increase of exponent $x$ with the field would lead to a reduction of the model oscillations (see Fig.~\ref{K_x_z}) and a somewhat improved fit to the SCGO data at 500~mT and 2~mT. This would imply that the correlation function of the system is slightly altered by the field.
\begin{table}
  \caption{Parameters entering the model for the interpretation of the data displayed in Fig.~\ref{data}.}\label{table}
  \begin{tabular}{l|ccc}
 & $x$ (-) & $f$ (-) & $(\Upsilon f)^{-1}$ ($\mu$s) \cr \hline
SrCr$_{8}$Ga$_{4}$O$_{19}$ & 0.25\,(5) & 0.038\,(6) & 0.095\,(5) \cr
ZnCu$_3$(OH)$_6$Cl$_2$ & 0.57\,(7) & 0.004\,(1) & 35\,(5)\cr
  \end{tabular}
\end{table}
As expected, the extracted values (Table~\ref{table}) for $x$ compare well with $\beta = 2-x $ of Refs.~\onlinecite{Uemura94,Mendels07}. An estimate for $\tau_{\rm e}$ can be derived from the fitted $\Upsilon$ values, although it crucially depends on $\Delta$. Concerning SCGO, for $\Delta$ = 0.5~T we obtain $\tau_{\rm e} \approx 10^{-11}$~s. For HS, with  $\Delta$ = 0.2~T we get $\tau_{\rm e} \approx 10^{-12}$~s. Obviously, these are preliminary estimates until the $\Delta$ values are determined.

{\sl Discussion and conclusions.}
SCGOO and HS are known for their imperfect structural properties. For instance HS is derived from atacamite Cu$_2$Cl(OH)$_3$ in which the Cu position forms a distorted pyrochlore lattice that can be viewed as a stacking of kagome and triangular planes. The substitution of Zn for Cu in the triangular plane of atacamite which leads to HS is not complete \cite{Freedman10}. The Cu spins remaining in the triangular planes are very weakly coupled to the kagome planes and essentially behave as free spins. Following the analysis of SCGO susceptibility data pointing to the coexistence of strongly and weakly coupled spins \cite{Schiffer97}, Henley \cite{Henley01} has noted that two adjacent vacancies on a triangle leave a lone spin with a field response suggesting spin fractionalization. The resulting ``half-orphan'' spin provides a reasonable fit \cite{Sen11,Sen12} to Ga NMR measurements \cite{Limot02}. 

Therefore the question arises whether the dynamics probed at different $\omega$ values  is associated with  quasi-free spins,  half-orphan spins, or with the intrinsic behavior of the kagome planes. For the last case we mention that correlations in the form of spin loops have been observed in the 3-dimensional analogue of the kagome lattice, i.e.\ the pyrochlore lattice \cite{Lee02}, and that the spectral response of spin chains with which the loops can be compared to in a coarse approximation is that of Eq.~\ref{Th_9} with $x = 0.5$ \cite{*[{See, e.g., Ref.~\onlinecite{vanHove54} or }][{}] Bloembergen49}. The underlying spin diffusion mechanism was suggested to explain the persistent spin dynamics often observed in geometrically frustrated magnets \cite{Yaouanc15}. This mechanism is however not expected to be at play in SCGO and HS since $F({\bf q},\omega)$ is ${\bf q}$-independent. It is however important to notice that extended spin correlations have been discovered through $\chi({\bf q})$ by neutron measurements in the millielectronvolt transfer energy range \cite{Han16,Nilsen13}. This has to be taken into account for a consistent discussion of the relaxation mechanism in that energy range.

For the discussion of our result and the comparison with published data, it is of interest to introduce the imaginary part of the dynamical susceptibility $\chi(\omega)$. This quantity is related to the spectral density through the fluctuation-dissipation theorem, which, in the high temperature limit $k_{\rm B}T/\hbar\omega \gg 1$ relevant to $\mu$SR measurements, reads ${\mathcal Im}\{\chi(\omega)\} \propto \omega F(\omega)$.

INS data recorded between $\approx$ 0.1 and 1~meV for deuterated herbertsmithite at low temperature lead to ${\mathcal Im}\{\chi(\omega)\} \propto \omega^\gamma$ with $\gamma = -0.7\,(3)$ \cite{Helton07,Helton10}. A crossover to some other regime is however expected since this diverging power-law cannot hold down to arbitrary low energies. Our result implies that ${\mathcal Im}\{\chi(\omega)\} \propto \omega^{x}$ with $x$ = 0.57\,(7) in the microelectronvolt energy scale. The question now is the physical significance of this result. An analysis of ${\mathcal Im}\{\chi(\omega)\}$ in the millielectronvolt energy range has been proposed in terms of the sum of contributions arising from defect spins and from the kagome planes \cite{Han16}. The latter contribution is negligible in the limit of small energies since the quoted analysis suggests a gapped spectrum of excitations in the kagome planes. The former contribution is modelled either as a damped harmonic oscillator or as a Lorentzian function \cite{Nilsen13,Han16}. While these functions are linear at small $\omega$, our result does not necessarily rule out defect spins as the origin of the $\mu$SR observations. In fact a promising route for further progress in our understanding of the kagome SL is the full determination of the spin correlations \cite{*[{For an example, see }][{}] Yaouanc20}. This requires the determination of the muon coupling parameters. In the case of insulators where the coupling is through the sole dipolar interaction, the problem amounts to the determination of the muon site. This can be achieved from experiments \cite{Amato97,Amato14,Dalmas16} or {\sl ab initio} calculations (see, e.g., Ref.~\onlinecite{Blundell22}).

We point out that half-integer spins are involved in SCGO and HS. Because of the well-known differences in the  magnetic response of spin chains involving half-integer and integer spins, it is of interest to study materials with the latter type of spins. Studies of the new kagome lattice system with $S = 1$  reported in Ref.~\onlinecite{Connolly18} could help in this direction.

For future progress it is probably a challenging but essential endeavor to compute $\Lambda({\bf q},\omega)$ for the diverse relaxation mechanisms mentioned above. Then the inference of the origin of the spin dynamics at different $\omega$ values should be within reach. On the experimental side further $\mu$SR field dependences measurements are required for a full characterization of the dynamics in the microelectronvolt range. Since large single crystals of HS are available, the study of the dependence of the relaxation on the orientation of the crystal is advisable. This would probe, for example, a possible deviation from the scalar nature of $\Lambda({\bf q},\omega)$.

We have discussed the long standing mystery of the $\mu$SR response in SL kagome systems. While the examination of the anomalous time scale is deferred to further work, we have targeted the question of the functional form of $P^{\rm exp}_Z(t)$. We provide a framework to consistently interpret the experimental data in terms of spin correlations decaying as a power-law of time. All parameters entering the model are susceptible of being computed from observables. A discussion of the energy dependence of the imaginary part of the dynamical susceptibility is provided. Further theoretical and experimental works are suggested for a better understanding of the spin dynamics in SL compounds, in particular to address the question of the gapped or gapless nature of the exitations in the kagome Heisenberg antiferromagnet. We note that understanding magnetic defects in frustrated two-dimensional systems is of interest in its own right \cite{Norman16} and may have implications for three-dimensional materials such as the pyrochlore compounds \cite{Patil20}.

We are grateful to A. Maisuradze for a critical reading of the manuscript.

\bibliography{reference}

\end{document}